# Rotating magnetic field driven antiferromagnetic domain wall motion: Role of Dzyaloshinskii-Moriya interaction

W. H. Li, Z. Y. Chen, D. L. Wen, D. Y. Chen, Z. Fan, M. Zeng, X. B. Lu, X. S. Gao, and M. H. Qin[*]

*Institute for Advanced Materials, South China Academy of Advanced Optoelectronics and Guangdong Provincial Key Laboratory of Quantum Engineering and Quantum Materials, South China Normal University, Guangzhou 510006, China*

**[Abstract]** In this work, we study the rotating magnetic field driven domain wall (DW) motion in antiferromagnetic nanowires, using the micromagnetic simulations of the classical Heisenberg spin model. We show that in low frequency region, the rotating field alone could efficiently drive the DW motion even in the absence of Dzyaloshinskii-Moriya interaction (DMI). In this case, the DW rotates synchronously with the magnetic field, and a stable precession torque is available and drives the DW motion with a steady velocity. In large frequency region, the DW only oscillates around its equilibrium position and cannot propagate. The dependences of the velocity and critical frequency differentiating the two motion modes on several parameters are investigated in details, and the direction of the DW motion can be controlled by modulating the initial phase of the field. Interestingly, a unidirectional DW motion is predicted attributing to the bulk DMI, and the nonzero velocity for high frequency is well explained. Thus, this work does provide useful information for further antiferromagnetic spintronics applications.

---

[*]Authors to whom correspondence should be addressed. Electronic mail: qinmh@scnu.edu.cn

**I. Introduction**

Antiferromagnetic (AFM) materials are attracting more and more attentions due to their potential applications in the newly emerged AFM spintronics field.[1-3] Specifically, zero net magnetization and ultralow susceptibility in AFM elements of future storage devices allow a significant improvement of performance stability against perturbing magnetic fields, as well as an enhancement of element density without any stray fields.[1,4] More importantly, magnetic dynamics much faster than in ferromagnets makes AFM spintronics more promising.[5,6] So far, several methods of modulating AFM domains and/or driving domain wall (DW) motion have been experimentally revealed or theoretically predicted.[7-9]

In experiments, it has been reported that the local staggered effective field in CuMnAs is induced by the applied electrical current through the spin-orbit effect, which controls the orientation of the AFM moments.[10] Similar behavior has been reported in $Mn_2Au$ whose magnetic atoms are also with the locally broken inversion symmetry.[11] More recently, the reversal of the magnetic domains in biaxial antiferromagnets NiO was realized by the current induced anti-damping torque.[12] Theoretically, the staggered field driven DW motion was investigated, and a high speed ~30 km/s was predicted partially attributing to the absence of Walker breakdown which normally limits the velocity of typical ferromagnetic DW.[8] Moreover, several external stimuli including spin wave excitations, field gradients, spin-orbit torques, and thermal gradients have been suggested to efficiently drive AFM DW motion.[13-23] For example, AFM DW motion under an applied temperature gradient has been revealed to be determined by the competition between the entropic torque and the Brownian force.[16]

On the other hand, the Dzyaloshinskii-Moriya interaction (DMI) may lead to faster and more controllable motion of AFM DW. For instance, the DMI induces the spin rotation symmetry breaking, resulting in the dependence of the DW motion on the polarization direction of the injected spin waves.[19,20] Furthermore, the rotating magnetic field driven chiral DW motion has been also predicted in antiferromagnets with DMI, providing broad opportunities of controlling DW.[24] It is suggested that the DMI plays an essential role in the motion, while the DW may not be efficiently driven by the rotating field alone. Specifically, in order to explain this phenomenon, the AFM DW is regarded as a combination of a head-to-head DW and a tail-to-tail DW (move in opposite directions under a rotating field for the ferromagnetic case) on which the driven torques are well cancelled.

However, the AFM dynamics could be much more complex due to the strong coupling between two magnetic sublattices, and the above explanation may not perfectly work. Specifically, when a rotating field **h**(*t*) is applied, the precession torque ~ –**S** × **H** and the damping torque ~ –**S** × (**S** × **H**) related to spin **S** and local effective field **H** are induced on the central plane of the AFM DW (detailed definition will be given in the next section), as depicted in Fig. 1 where shows a Néel AFM DW in a nanowire without DMI. It is well known that for an AFM system, applied field **h** tends to align the spins in perpendicular to **h** to save the Zeeman energy,[25] and the DW is expected to rotate synchronously with the rotating **h**. As a result, a stable precession torque is available, resulting in an efficient DW motion. As a matter of fact, comparing with strong AFM exchange coupling, DMI may be too weak to be considered in a considerable number of antiferromagnets. Thus, there is still an urgent need to make clear the role of DMI in the rotating field driven AFM DW motion.

In this work, we study the AFM DW motion driven by the rotating field using the Landau-Lifshitz-Gilbert (LLG) simulations of the classical Heisenberg spin model. We figure out that even in the absence of DMI, the rotating field at a low frequency could efficiently drive the DW motion with a steady velocity, while in large frequency region, the DW only oscillates around its equilibrium position and cannot propagate. The dependences of the velocity and the critical frequency on several parameters are investigated and discussed in detail. More interestingly, in the presence of DMI, a unidirectional DW motion driven by the rotating field is predicted and well explained.[26]

## II. Model and method

We start from an AFM spin model with isotropic Heisenberg exchanges between the nearest neighbors and a uniaxial anisotropy term

$$H = J \sum_{<i,j>} \mathbf{S}_i \cdot \mathbf{S}_j - d_z \sum_i \left(S_i^z\right)^2 - \sum_i \mathbf{h}(t) \cdot \mathbf{S}_i, \tag{1}$$

where $J > 0$ is the AFM coupling constant, $d_z > 0$ is the anisotropy constant defining an easy axis in the $z$ (nanowire axis) direction, $\mathbf{S}_i = \boldsymbol{\mu}_i/\mu_s$ represents the normalized magnetic moment at site $i$ with the three components $S_i^x$, $S_i^y$ and $S_i^z$, $\mathbf{h}(t)$ is the circularly polarized magnetic field $\mathbf{h}(t) = h_0(\cos\omega t \cdot \mathbf{e}_x + \sin\omega t \cdot \mathbf{e}_y)$ with frequency $\omega$ and amplitude $h_0$ applied in the $xy$ plane, as shown in Fig. 1. Here, the model is chosen similar to the earlier work, and the Néel AFM DW is investigated,[24] allowing one to understand the role of the DMI easily.

Subsequently, the spin dynamics is investigated by the stochastic LLG equation,[27-29]

$$\frac{\partial \mathbf{S}_i}{\partial t} = -\frac{\gamma}{\mu_s(1+\alpha^2)} \mathbf{S}_i \times \left[\mathbf{H}_i + \alpha(\mathbf{S}_i \times \mathbf{H}_i)\right], \tag{2}$$

where $\gamma$ is the gyromagnetic ratio, $\alpha$ is the Gilbert damping constant, and $\mathbf{H}_i = -\partial H/\partial \mathbf{S}_i$ is the effective field. Unless stated elsewhere, the LLG simulations are performed on an one-dimensional lattice (lattice parameter $a$) with $1\times1\times1001$ spins with open boundary conditions using fourth-order Runge-Kutta method with a time step $\Delta t = 2.0 \times 10^{-4} \mu_s/\gamma J$, $d_z = 0.01J$, $\alpha = 0.01$, and $\gamma = 1$.[30] After sufficient relaxation of the AFM DW, $\mathbf{h}(t)$ is applied and the DW motion is studied.

## III. Simulation results and discussion

*3.1 Rotating field driven domain wall motion in the absence of DMI*

Fig. 2 presents the DW position as a function of time for various $\omega$ under $h_0 = 0.1J$, which reveals two driven motion modes divided by a critical frequency $\omega_c$. For a small $\omega$ below $\omega_c$, for example $\omega = 0.05\gamma J/\mu_s$, a stable DW propagation with a constant velocity is observed, demonstrating that the rotating field alone can drive the AFM DW motion even in the absence of DMI. Similar to the chiral DW motion, the drift velocity increases with $\omega$, as shown in Fig. 2(a). When $\omega$ is increased above $\omega_c \sim 0.098\gamma J/\mu_s$, the rotating field could not efficiently drive the DW anymore, and the DW only oscillates around its equilibrium position ($z = 505a$), as shown in Fig. 2(b) where gives the $t$-dependent DW position for various $\omega$ above $\omega_c$.

The velocity $v$ as a function of $\omega$ for $h_0 = 0.1J$ is summarized in Fig. 2(c), and the simulated results can be qualitatively understood from the competing torques acting on the DW.[15] When $\mathbf{h} = h_0(0, 1, 0)$ is applied, for instance, the precession torque $\mathbf{\Gamma}_{p,h} \sim -\mathbf{S} \times \mathbf{h}$ and the damping torque $\mathbf{\Gamma}_{d,h} \sim -\mathbf{S} \times (\mathbf{S} \times \mathbf{h})$ are induced, as depicted in Fig. 2(d). Subsequently, the spins are driven slightly out of the easy plane, resulting in the additional precession torque $\mathbf{\Gamma}_{p,ex} \sim -\mathbf{S} \times \mathbf{H}_{ex}$ due to the strong AFM exchange interaction, where $\mathbf{H}_{ex}$ is the exchanged field. It is noted that $\mathbf{\Gamma}_{p,ex}$ is opposite to $\mathbf{\Gamma}_{p,h}$ and quickly suppresses the total precession torque $\sim -\mathbf{S} \times \mathbf{H}$. As a result, a static field alone could not drive the DW motion, as has been explained in detail in our earlier work.[31] However, the effect of the rotating field on AFM dynamics is rather different. On one hand, the rotating field $\mathbf{h}$ tend to align the DW spins in perpendicular to $\mathbf{h}$ to save Zeeman energy. However, the spin rotation is slightly behind $\mathbf{h}$ (schematically depicted by $\Delta\delta$

= δ – δ' for blue spin in Fig. 2(d)), leading to that $\Gamma_{p,h}$ always defeat $\Gamma_{p,ex}$ and in turn resulting in a stable DW motion. Simultaneously, $\Gamma_{d,h}$ on the up spin is larger than that on the down spin. Thus, a net damping torque on the DW is generated and drive the DW plane synchronously rotates with **h**.

This argument has been confirmed by tracing the DW profile and tilt angle of the DW plane $\phi$, as respectively shown in Figs. 3(a) and 3(b). The normalized staggered magnetization **n** = ($S_{2i}$ – $S_{2i-1}$)/| $S_{2i}$ – $S_{2i-1}$| for $\omega = 0.06\gamma J/\mu_s$ in Fig. 3(a) shows that $n_x$ and $n_y$ of the DW spins evolve with time, demonstrating the rotation of the DW plane. Moreover, $\phi$ linearly increases with $t$, as shown in Fig. 3(b), clearly demonstrating the synchronized precession of the DW with the rotating field with a fixed phase difference ($\omega t - \phi$ approximately equals to δ). When $\omega$ is increased, the net damping torque should be enhanced resulted from the increase of Δδ to drive the wall plane catch up with the rotating field. Similarly, the net precession torque $\Gamma_{p,h} - \Gamma_{p,ex}$ is also enhanced, resulting in the increase of $v$. However, when $\omega$ increases above the critical value, the DW spins cannot catch up with **h** anymore ($\omega t - \phi$ changes with time for $\omega = 0.11\gamma J/\mu_s$ in Fig. 3(b), for example), and a periodically oscillating $\Gamma_{p,h}$ is induced. Thus, the DW only oscillates around its equilibrium position and cannot propagate.

*3.2 Effects of external and internal fields*

Subsequently, we investigate the dependences of $v$ and $\omega_c$ on various parameters. Fig. 4(a) gives the simulated $v$ as a function of $\omega$ for various $h_0$, which clearly shows that both $v$ and $\omega_c$ increase with the increasing $h_0$. It is noted that the net precession torque is enhanced with $h_0$, resulting in the increase of $v$ for a fixed $\omega$. Furthermore, the net damping torque on the DW is also increased, which in turn significantly enhances the rotation capability of the DW, resulting in the increase of $\omega_c$. Taking NiO as an example to estimate the real physical values, we set the exchange stiffness $A \approx 5 \times 10^{-13}$ J m$^{-1}$, lattice constant $a \approx 4.2$ Å, and magnetic moment $\mu_s \approx 1.7\mu_B$. For $h = 0.1$ T and $d_z = 0.01$, the critical frequency and velocity are estimated to be $\omega/2\pi \sim 373$GHz and $v \sim 100$ms$^{-1}$, respectively. As a matter of fact, a weak $h_0 \sim 1$ mT can also drive the DW motion, although the critical frequency and velocity are rather small.

Moreover, the mobility of the DW is significantly dependent of the DW energy $E_{DW}$,[31] i. e., higher $E_{DW}$ leads to lower mobility. This property is confirmed in our simulations, as shown in Fig. 4(b) where gives the $v(\omega)$ curves for various lattice dimensions. It is noted that $E_{DW}$ is extensively increased with the increase of the lattice dimension, resulting in the decrease of $v$ for a fixed $\omega$. Specifically, the exchange field is significantly enhanced, and the rotation capability of the DW plane is suppressed. As a result, the critical frequency $\omega_c$ decreases with the increasing lattice dimension, similar to the earlier report.[24] For a fixed lattice dimension, the finite-lattice size effect has been checked to be negligible and our conclusion is reliable, although the corresponding results are not shown here.

Fig. 4(c) presents the simulated $v(\omega)$ curves for various $d_z$. Based on the continuum model, the DW energy can be estimated by $E_{DW} = 2(2|J|d_z)^{1/2}$.[32] With the increase of $d_z$, the DW energy is increased, resulting in the decrease of $v$ for a fixed $\omega$. Furthermore, the damping torque resulted from the exchange interaction $\Gamma_{d,ex}$ is slightly suppressed and the net damping torque is strengthened, resulting in the enhancement of the DW rotation capability. In other words, the enhanced $d_z$ could speed up the rotation of the DW plane, allowing the DW catch up with **h** easily and resulting in the increase of $\omega_c$. Similarly, the effect of the intermediate anisotropy energy $d_x\Sigma_i(S_i^x)^2$ available in some AFM materials on the DW motion is also investigated, and the corresponding results are given in Fig. 4(d). It is noted that the additional anisotropic energy barrier should be conquered during the rotation of the DW when a nonzero $d_x$ is considered, resulting in the suppression of the DW rotation capability. Specifically, the introduced $d_x$ breaks the rotation symmetry of the DW, seriously impeding the synchronous rotation of the DW with **h**. Thus, $\omega_c$ is significantly decreased with the increase of $d_x$. In addition, $\Gamma_{p,ex}$ is slightly enhanced, and $v$ slowly decreases as $d_x$ increases.

Furthermore, it is noted that the damping term always impedes the spin precession and suppresses the DW mobility and rotation capability, resulting in the decreases of $v$ and $\omega_c$, as reported in the earlier works.[31,33] This behavior has been checked in our simulations, and the corresponding results are not shown here for brevity.

*3.3 Roles of DMI*

So far, the rotating field driven AFM DW in the absence of DMI has been clearly demonstrated in our LLG simulations. In this part, we check the dependence of the DW dynamics on DMI and compare the results with the earlier report.

The simulated DW velocity under various field strength for a small DMI constant $D = 0.002J$ with $\alpha = 0.005$ are presented (solid points) in Fig. 5(a) where the analytical results obtained from the earlier derivation are also given by solid lines.[24] The simulation results for $h_0 = 0.02J$ agree well with the theoretical solution, well confirming the validity of our simulation results. However, the simulated results remarkably deviate from the analytical results below $\omega_c$ for a larger field $h_0 = 0.03J$, indicating novel physics could be available. Moreover, the fixed phase difference between the synchronized motion and the rotating field could be modulated by tuning the initial phase of the rotating field, and controls the bidirectional motion (backward or forward) of DW for $w < w_c$. For example, in the absence of DMI, $v(h_0) = -v(-h_0)$ is clearly shown in Fig. 5(b) where presents the simulated $v$ as a function of the field magnitude $|h_0|$. More importantly, for $D = 0.01J$, the DW propagates along the same direction with $0 < v(h_0) < v(-h_0)$, similar to the earlier report.[26] Thus, our work demonstrates again that an explicit space-time asymmetry is not essential for realizing the DW unidirectional motion. This phenomenon is qualitatively explained below to help one to understand the physics better.

Through the symmetry analysis, one obtains the general constraint:

$$v(D, h_0) = -v(-D, -h_0). \tag{3}$$

Following the earlier work,[26] the velocity is perturbatively expanded to be:

$$v(D, h_0) = v_0(h_0) + Dv_1(h_0) + D^2v_2(h_0) + \cdots, \tag{4}$$

with $v_n(h_0) = (-1)^{n+1}v_n(-h_0)$ to satisfy the general constraint condition. The equation clearly shows that the odd (even) order DMI contributions to the velocity are unidirectional (bidirectional). We calculate the odd order contributions $v_{odd} = [v(D, -h_0) + v(D, h_0)]/2 = Dv_1(h_0) + D^3v_3(h_0) + \cdots$, and give the results in Fig. 5(c). On one hand, the first order contribution $Dv_1$ is larger than the zeroth order contribution $v_0$ for $\omega > \omega_c$ and $D > 0.005J$, due to the fact that $Dv_1$ comes from an additional unidirectional torque induced by the domain wall chirality.[26,34] Moreover, $v_{odd}$ are unidirectional for finite $D$, resulting in the effective DW motion even for $\omega > \omega_c$, noting that the even order contribution is compensated in an oscillation period.

On the other hand, with the increase of $|h_0|$, $|v_0|$ significantly increases while $v_{odd}$ is almost unchanged. As a result, for small $|h_0|/D$, the DW velocity is dominated by the DMI, and a unidirectional DW motion is observed. However, with the increase of $|h_0|$ or decrease of $D$, the DMI effect on the velocity is effectively suppressed, and the even order contribution is significantly enhanced for $\omega < \omega_c$. As a matter of fact, the DMI effect seems to be overemphasized in the earlier work due to the ignorance of $v_0$.[24] In Fig. 5(d), the calculated $v_{odd}/D$ as a function of $|h_0|$ for various $D$ are presented, which shows the independence of $v_{odd}/D$ on $D$, demonstrating the validity of the perturbative expansion in Eq. (4). Moreover, the calculated $v_{even} - v_0 = [v(D, -h_0) + v(D, h_0)]/2 - v_0 = D^2v_2(h_0) + D^4v_4(h_0) + \cdots$ is two orders of magnitude smaller than $v_0$ (not shown here), demonstrating that the high order (n > 2) can be safely neglected and the above analysis is reliable.

Thus, the large deviation between simulation and analytical theory for $\omega < \omega_c$ under large $h_0$ comes from the zeroth order contribution which cannot be simply neglected. Since the zeroth order contribution is bidirectional and is compensated in one period for $\omega > \omega_c$, resulting in the well consistency of the simulation and theory above $\omega_c$ even under a large $h_0$.

**IV. Conclusion**

In summary, we have studied the rotating field driven AFM DW motion using the LLG simulations of the classical Heisenberg spin model. It is revealed that even in the absence of the DMI, the rotating field in the low frequency below $\omega_c$ could efficiently drive the DW motion. In this case, the DW rotates synchronously with the rotational field, and a stable precession torque is available and drives the DW motion with a steady velocity. In the large frequency region, the DW only oscillates around its equilibrium position and cannot propagate. Meanwhile, the dependences of the velocity and critical frequency on several parameters have been investigated in detail and qualitatively explained. Interestingly, the initial phase of the rotating field can control the bidirectional motion of DW without DMI, while the DMI could lead to the unidirectional DW motion.

**Acknowledgment**

We are grateful for insightful discussions with Huaiyang Yuan, and Zhengren Yan. The work is supported by the National Key Projects for Basic Research of China (Grant No. 2015CB921202), and the Natural Science Foundation of China (No. 51971096), and the Science and Technology Planning Project of Guangzhou in China (Grant No. 201904010019), and the Natural Science Foundation of Guangdong Province (Grant No. 2016A030308019).

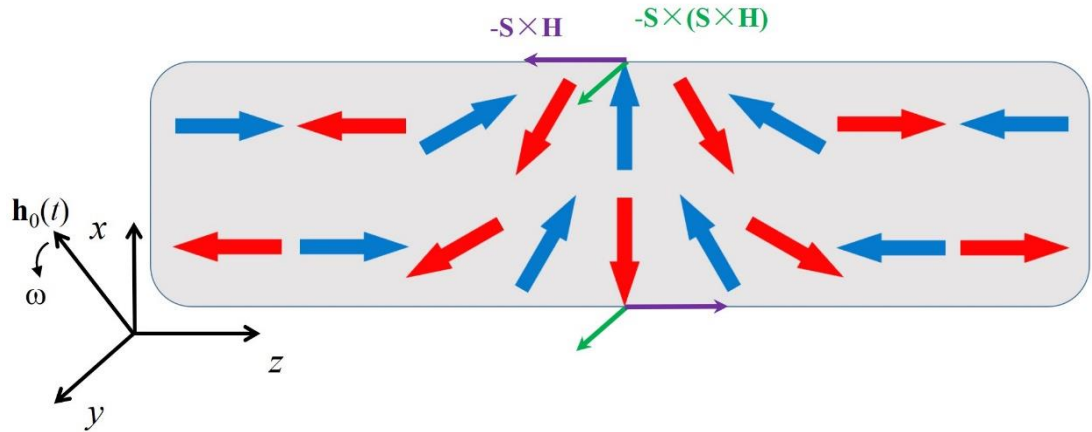

Fig.1. (color online) Sketch of the torques acting on the central plane of a domain wall under a rotating magnetic field. The two sublattices of the AFM are occupied with blue and red spins, respectively. The rotating field in the *xy* plane is also depicted.

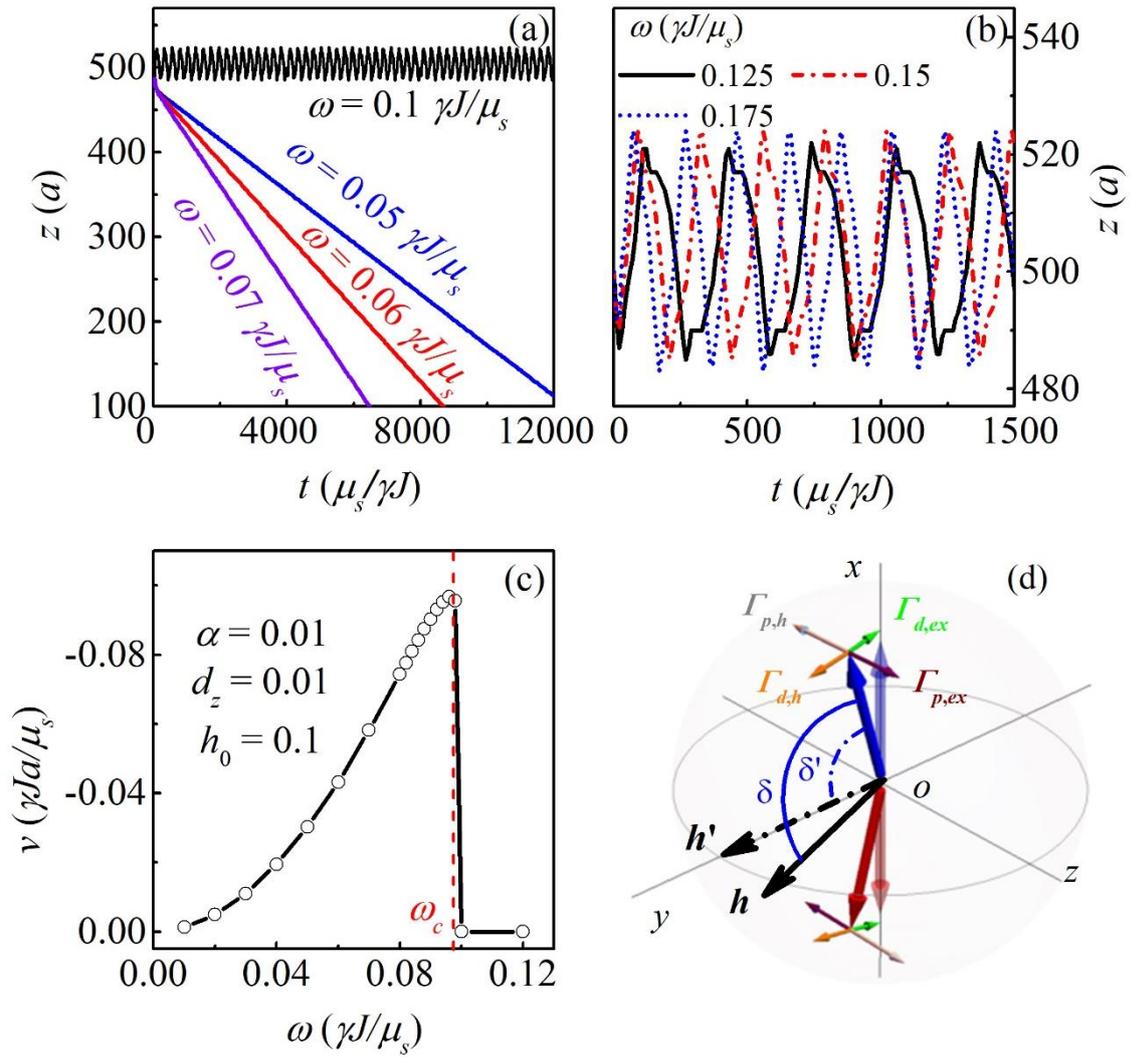

Fig.2. (color online) The domain wall position as a function of time $t$ for (a) $\omega$ = 0.05, 0.06, 0.07 and 0.1$\gamma J/\mu_s$, and (b) $\omega$ = 0.125, 0.15 and 0.175$\gamma J/\mu_s$ at $h_0$ = 0.10. (c) The calculated DW drift velocity $v$ as a function of $\omega$ under $h_0$ = 0.1, and (d) a schematic depiction of the torques acting on the central plane of a DW under a rotating field.

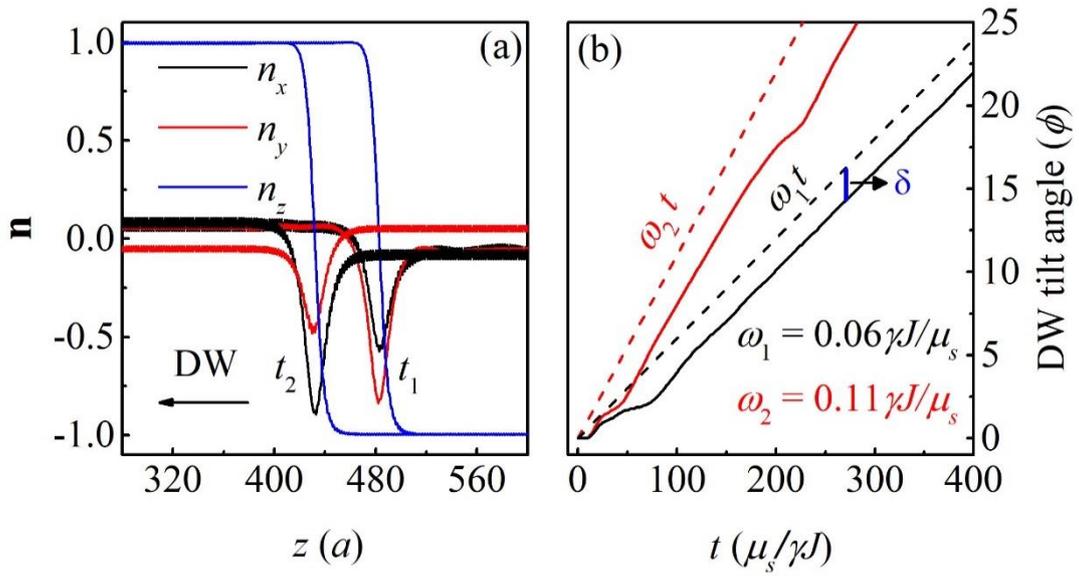

Fig.3. (color online) (a) The domain wall profiles for $\omega = 0.06\gamma J/\mu_s$ at $t_1$ and $t_2$, and (b) the DW tilt angle as a function of time for $\omega_1 = 0.06\gamma J/\mu_s$ and $\omega_2 = 0.11\gamma J/\mu_s$. The phases $\omega_1 t$ and $\omega_2 t$ are also given for an easy comparison.

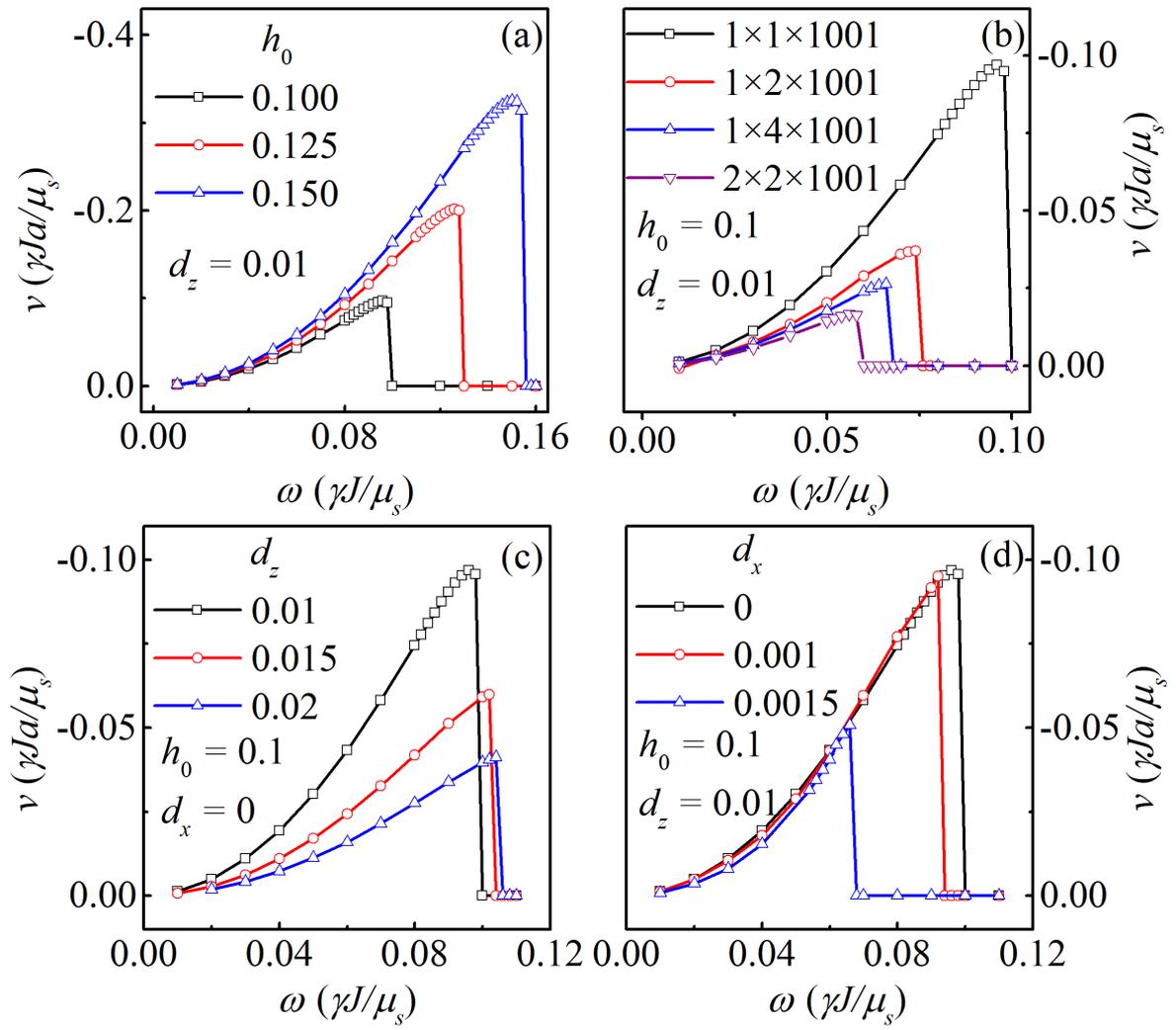

Fig.4. (color online) The calculated $v(\omega)$ curves (a) for various $h_0$, and (b) for various lattice sizes, and (c) for various $d_z$, and (d) for various $d_x$ for $d_z = 0.01$.

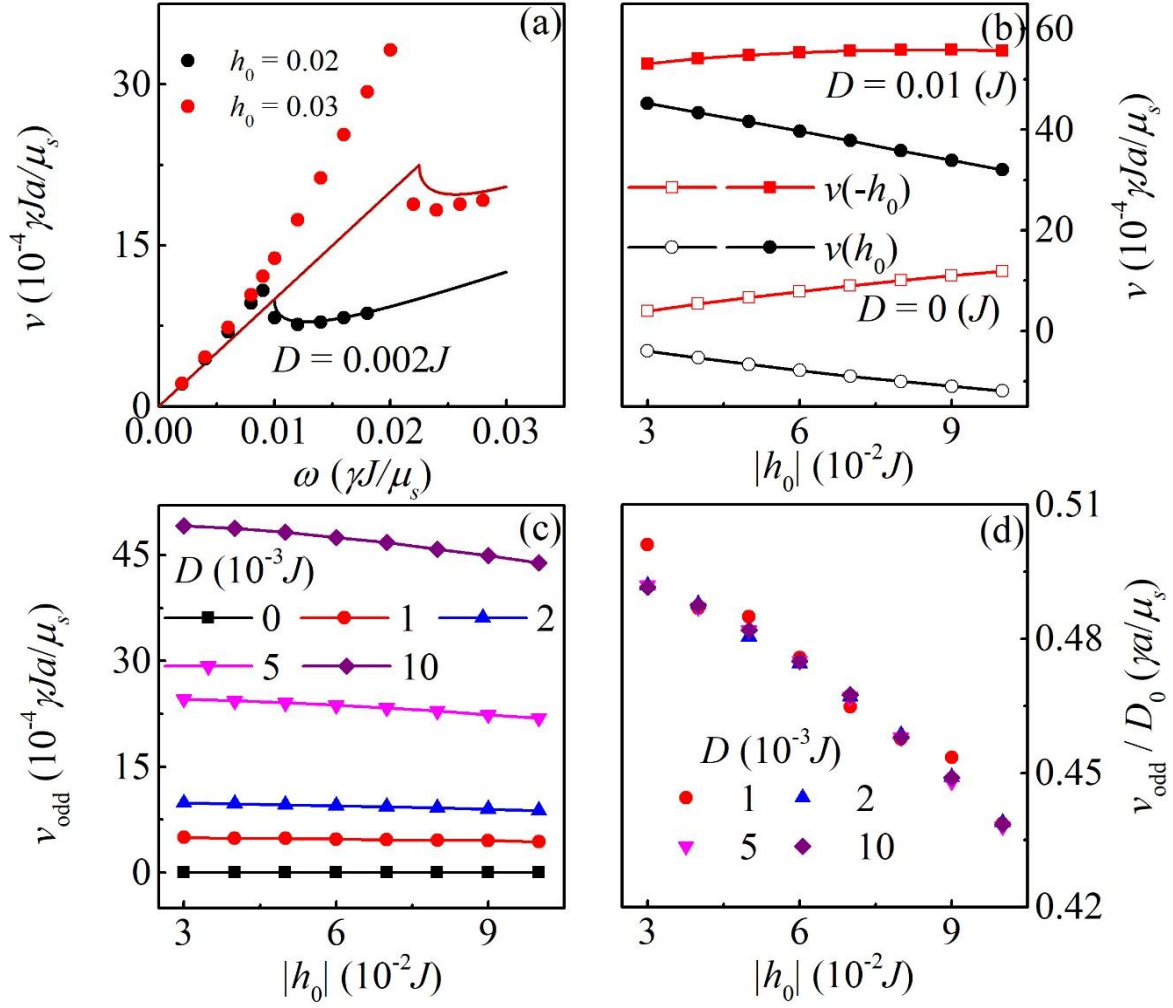

Fig.5. (color online) (a) The simulated $v(\omega)$ (solid circles) with various $h_0$ for $D = 0.002J$ and $\alpha = 0.005$. The solid lines are plotted using the analytical equations (12) and (17) in Ref. [24]. (b) The calculated $v(\pm h_0)$ for $D = 0$ and $D = 0.01(J)$ for $\omega = 0.01\gamma J/\mu_s$, and (c) $v_{odd}$, and (d) $v_{odd}/D$ as functions of $h_0$ for various $D$.